\documentclass[%
 preprint, amsmath, amssymb,
 aps, prfluids, superscriptaddress
]{revtex4-2}

\usepackage{graphicx}
\usepackage{dcolumn}
\usepackage{hyperref}
\usepackage{xcolor}

\newcommand{\pco}{$P_{\text{CO}_{2}}$}
\begin{document}

\title{Experimental and numerical study of CO$_{2}$ dissolution in a heterogeneous Hele-Shaw cell}
\author{Rima Benhammadi}
\affiliation{Institute of Environmental Assessment and Water Research, Spanish National Research Council (IDAEA-CSIC), Barcelona, Spain}
\author{Patrice Meunier}
\affiliation{Institut de Recherche sur les Ph\'enom\`enes Hors \'Equilibre (IRPHE), Marseille, France}
\author{Juan J. Hidalgo}
\email{juanj.hidalgo@idaea.csic.es}
\affiliation{Institute of Environmental Assessment and Water Research, Spanish National Research Council (IDAEA-CSIC), Barcelona, Spain}
\date{\today}
%
%
\begin{abstract}
  We investigate the convective instability resulting from the dissolution of carbon dioxide (CO$_{2}$) into water in a heterogeneous Hele-Shaw cell utilizing both experimental and numerical approaches. Experiments are conducted in a Hele-Shaw cell with a variable gap width corresponding to a log-normally distributed permeability of variance $\sigma_{\log K}^2 = 0.135$. Two mean gaps (370 $\mu$m and 500 $\mu$m) with the same correlation lengths ($\lambda_x$ = 0.032 m and $\lambda_z$ = 0.016 m) are considered. Experiments in homogeneous cells with a constant gap are also performed. The CO$_2$ partial pressure (\pco) is varied between $12\%\pm 1\%$ (0.12 bar) and $85\%\pm 1\%$ (0.85 bar). The convective patterns are visualized using Bromocresol green. The effect of the heterogeneity on the instability is analyzed through its wavenumber, amplitude and growth rate. There is a good agreement between the experimental and numerical results. Fingers appear more dispersive and distorted in the heterogeneous media. Heterogeneous cases display a larger instability amplitude, faster growth rate and smaller dimensionless wavenumber. This reflects that heterogeneity accelerates the instability and the merging of the fingers. A comparison of the autocorrelation function of the fingering patterns and the permeability field shows that heterogeneity increases the dimensionless correlation length of the fingering pattern, which slows down its growth when its size becomes comparable to the heterogeneity. 
\end{abstract}
%
%
%
\maketitle
%
%
\section{Introduction}
The need to reduce anthropogenic greenhouse gas emissions has catalyzed extensive research into carbon dioxide (CO$_{2}$) sequestration, a technology aimed at reducing the environmental impact of CO$_{2}$. Among the various options for sequestration, injecting CO$_{2}$ into saline aquifers stands out due to its significant storage capacity, surpassing that of depleted hydrocarbon reservoirs \cite{Ershadnia2021}. This intrinsic advantage has driven a surge in scientific inquiry and technological innovations over the past few decades \cite{Ajayi2019}. Numerous studies have investigated CO$_{2}$ sequestration in homogeneous porous media experimentally using analogue fluids \cite{article_Emami, ching_convective_2017} and in 3D granular porous media \cite{article_Brouzet}. A comprehensive review of experimental, modeling, and field studies of CO$_{2}$ sequestration in various underground formations can be found in Kalam et al. (2020) \cite{Kalam2020}. However, research specifically addressing the implications of heterogeneity on convective mixing remains limited.

The effect of heterogeneity has been mainly studied numerically. Heterogeneity reduces onset times \cite{article_Green}, notably when permeability increases \cite{chen2013}. However, in vertically stratified media, the onset of perturbation can be delayed or accelerated depending on the relationship between the wavenumber of the permeability variation and the specific perturbation mode \cite{Ghorbani2017}.
Heterogeneity alters the concentration patterns during CO$_{2}$ dissolution. Depending on the correlation length and variance of the permeability, three flow regimes can be identified: fingering, dispersive, and channeling \cite{article_Farajzadeh, ranganathan2012}. Numerical simulations also showed that heterogeneity induces more pronounced fingering competition, greater variability in finger width, and a reduced mixing interfacial length \cite{Li2019}.


Research has also shown that heterogeneity impacts CO$_{2}$ trapping and dissolution processes. De Paoli et al. (2016) \cite{DePaoli2016} demonstrated that anisotropic sedimentary rocks can dissolve significantly more CO$_{2}$ than isotropic rocks. The main factor affecting dissolution fluxes is permeability connectivity, with highly connected structures being more efficient at trapping CO$_{2}$ \cite{Ershadnia2021} and channeling \cite{zhang2024} increasing mass transfer. However, the correlation length of the permeability has a limited effect on the dissolution flux \cite{ranganathan2012}.

Previous experimental studies have primarily focused on homogeneous media \cite{KneafPru2011, Backhaus2011, Faisal2015, article_Outeda, article_vreme}. Few experiments have been conducted in heterogeneous porous media. Trevisan et al. (2015) performed experiments in synthetic sand reservoirs to investigate the impact of capillary heterogeneity on flow dynamics and the trapping efficiency of supercritical CO$_{2}$ \cite{Trevisan2015}. Furthermore, Fern\o\ et al. (2023) \cite{ferno2023} conducted repeated CO$_{2}$ injections in a physical model of a faulted geological cross-section, revealing that heterogeneity reduces reproducibility, especially in fault-related areas. Notably, there has been no exploration of heterogeneous Hele-Shaw cells.

Our study hence aims to understand the impact of permeability heterogeneity on the convective instability generated during CO$_{2}$ dissolution into water by analyzing key parameters, such as the instability wavenumber, amplitude and growth rate. We restrict our analysis to the Darcy regime, ensuring consistency with experimental and numerical setups. Methodologically, our approach integrates experiments conducted in two-dimensional Hele-Shaw cells with varying and constant gap thicknesses, alongside numerical simulations. The manuscript is organized as follows: First, we describe the experimental and numerical methodology. Then, we detail the experimental and numerical results and analyze the structure of the convective instability and fingering patterns. At the end, some concluding remarks are provided.
%
%
\section{Methodology}
We study the dissolution of CO$_{2}$ in water and the associated convective instability in homogeneous and heterogeneous media by means of experimental and numerical observations. In the following we describe the experimental set up, the experimental images analysis and the numerical methods used in the simulations.

\subsection{Experimental setup}
\begin{figure}
    \centering
    \includegraphics[width=\textwidth]{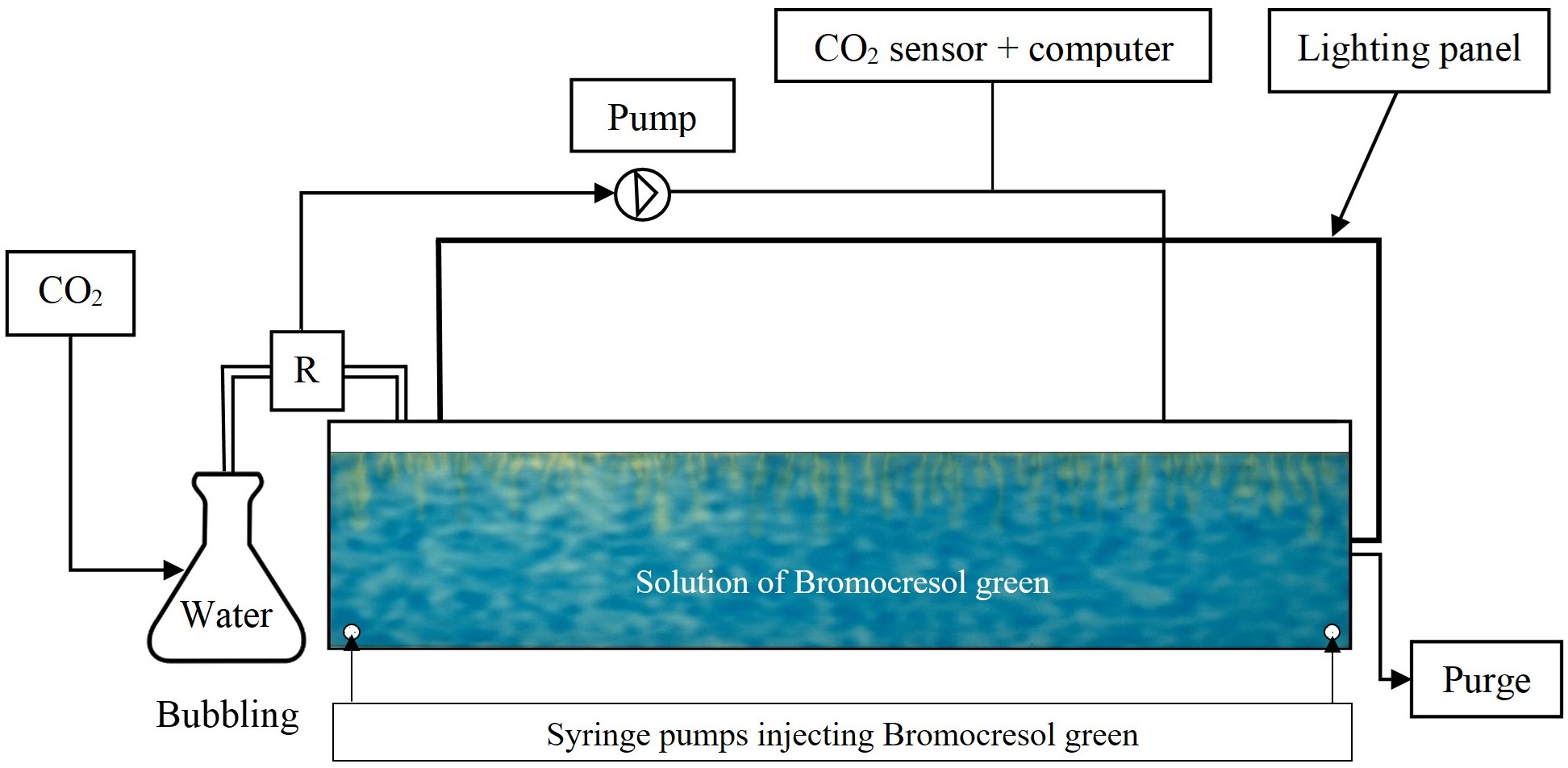}
    \caption{Experimental setup. The Hele-Shaw is 0.49 m wide and it is filled up to a height of 0.073 m with a solution of Bromocresol green, a pH sensitive dye, used to visualize the fingering patterns. The CO$_{2}$ is initially bubbled in order to reduce its evaporation then it is injected in the cell. A lighting panel is placed behind the cell and the purge is kept opened during the injection of CO$_{2}$ to allow the air to escape and thus avoid the interface from dropping drastically.} \label{fig:exp_setup}
\end{figure}

A schematic of the experiment is presented in Figure~\ref{fig:exp_setup}. The Hele-Shaw cells were constructed using polymethylmethacrylate. They have a length  $L=0.49$ m. The cell is initially filled up to a height $H = 0.073$ m with a $5 \times 10^{-4}$ mol/L aqueous solution of basic Bromocresol green, a pH-sensitive dye. When CO$_{2}$ is introduced above the free surface of the aqueous solution, it dissolves in a thin diffusive layer just below the interface. The high density of the water enriched by CO$_{2}$ triggers the convective instability. Since the CO$_{2}$ is acidic, the color of the front turns yellow when dissolution starts. CO$_{2}$ is bubbled in an Erlenmeyer flask filled with 250 ml of water to minimize evaporation and then injected into another small reservoir (R in Figure~\ref{fig:exp_setup}) connected to the Hele-Shaw cell by a pump. During injection a purge remains open to allow air to escape and prevent a drastic drop in the interface level. When the desired CO$_{2}$ pressure, \pco, is reached, the injection is stopped and the purge closed. Experiments were performed for $P_{\text{CO}_{2}} =85\%, 74\%, 60\%, 46\%, 31\%, 20\%$ and $12\%$. A lighting panel is placed behind the cell and a camera in front. The field of view of the camera excluded the left 3 cm of the cell. Images were acquired every 30 seconds.
%
%
%

Experiments were conducted in homogeneous cells with constant gap width in order to compare the experimental observations of the heterogeneous cases. For the homogeneous cells, gap widths $E = 370, 500$~$\mu$m were chosen. These values ensure that the Brinkman term is negligible \cite{article_vreme}, i.e., $Ra\sqrt{Da}<10$, where
\begin{align}
  \label{eq:Ra}
  Ra = \frac{u_{b}H}{\phi D}
\end{align}
\begin{align}
  \label{eq:Da}
  Da = \frac{k}{H^2}
\end{align}
with $u_{b}=\frac{k \Delta \rho g}{\mu}$ is the buoyancy velocity, $D = 1.76\times10^{-9}$ m$^2$/s is the diffusion coefficient of CO$_{2}$ in water, $\phi$ the porosity (equal to 1 here), $k=\frac{E^2}{12}$ is the cell permeability and $g$ is the gravitational acceleration. The density difference $\Delta \rho$ depends on the CO$_{2}$ partial pressure as
\begin{align}
  \label{eq:delta_rho}
  \Delta \rho = \rho_{0}\alpha k_{H} P_{\text{CO}_{2}},
\end{align}
where $\rho_{0}$ is the density of water, $\alpha=\frac{1}{\rho_{0}} \frac{\partial \rho}{\partial c}$ is the chemical expansion coefficient of the density, $k_{H}$ the Henry’s constant of CO$_{2}$ in water ($\alpha k_{H}=0.004\pm 0.001$ MPa$^{-1}$), and $\mu=0.001$ kg/(m$\cdot$s) is the viscosity of water considered constant \cite{McBrideWright2014}. The parameters of the different experiments are summarized in Table~\ref{tab:params}.

Heterogeneous cells were constructed by engraving one side of the cell with a gap thickness pattern with an accuracy of $\pm 50 \mu $m. The pattern was obtained by transforming to depth a log-normally distributed permeability field generated using a Gaussian correlation function with horizontal correlation length $\lambda_{x} = 0.032$ m and vertical correlation length $\lambda_z = \lambda_{x}/2$, to obtain a permeability structure resembling the natural stratification of sedimentary CO$_{2}$ reservoirs. The selected $\lambda_{x}$ is equal to ten times the theoretical critical wavelength proposed by Riaz \emph{et al.}, 2006 \cite{riaz_2006} $\lambda_{c} = 2 \pi H/( 0.07 Ra)$ with $Ra = 2034$, which corresponds to $P_{\text{CO}_{2}}=60\%$. This decision was made to ensure that the size of the heterogeneity exceeds the scale of the instability. The variance of the log-permeability field was chosen as $\sigma_{\log K}^2 = 0.135$. this value ensures a permeability variation as large as possible while keeping the thickness nearly uniform in order for the flow to remain 2D at first order. Before engraving, the depth distribution was truncated to avoid very large and very small depth values.

\begin{figure}
    \centering
    \includegraphics[width=\textwidth]{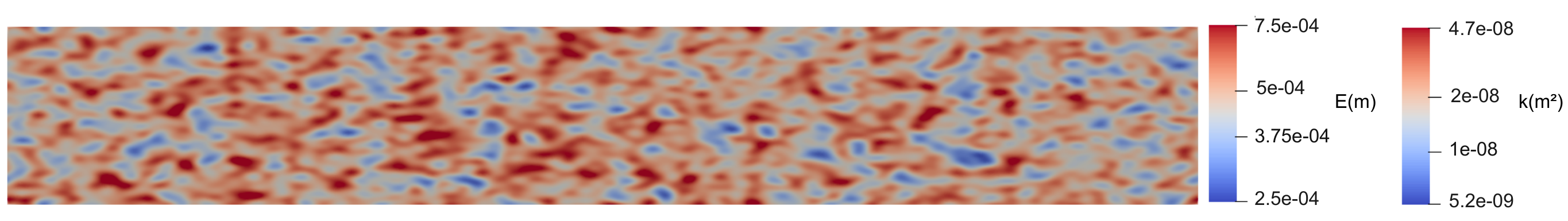}
    \caption{Engraved gap depth $E$ and permeability $k$ for the case $E_{1}$ with 250 $\mu$m $< E < 750$ $\mu$m.  Note the logarithmic color scale.}\label{fig:Kfield}
  \end{figure}

The mean permeability was changed by adjusting the distance between the two plates of the cell. Two thickness ranges were used: 250 $\mu$m $< E < 750$ $\mu$m (Figure~\ref{fig:Kfield}) and  120 $\mu$m $< E < 620$ $\mu$m. Both have a variance $\sigma_{E}=8.42 \times 10^{-9}$ m$^2$.
\begin{table}
    \centering
\caption{Values of $\Delta \rho$, $Ra$ and $Pe$ for the different gaps and \pco\ considered. The case $E = 370 \mu$m has a permeability $k = 1.14e-8$ m$^{2}$ and the case $E = 500 \mu$m, $k = 2.08e-8$ m$^{2}$.} \label{tab:params}
\begin{tabular}{|cc|cc|cc|}
  \cline{3-6}
 \multicolumn{2}{c}{}  &\multicolumn{2}{|c|}{ $E = 370 \mu$m} &\multicolumn{2}{c|}{ $E = 500 \mu$m}\\ \hline
 \pco\ (\%) & $\Delta \rho$ (kg/m$^{3}$) &  $Ra$ & $Pe$ & $Ra$  & $Pe$\\ \hline
  12 & 0.048 & 407  & 2.8   & 222.8  & 1.1  \\
  20 & 0.08  & 678  & 4.6   & 371.4  & 1.9  \\
  31 & 0.124 & 1051 & 7.2   & 575.6  & 2.9  \\
  46 & 0.184 & 1560 & 10.7  & 854.1  & 4.33 \\
  74 & 0.304 & 2577 & 17.7  & 1411.1 & 7.2  \\
  85 & 0.34  & 2882 & 19.7  & 1578.3 & 8.0  \\  
  \hline
\end{tabular}
\end{table}

%
%
\subsection{Processing of the experimental images}
\begin{figure}
    \centering
    \includegraphics[width=14cm]{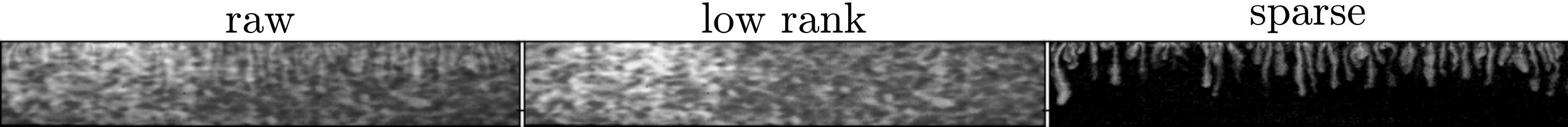}
    \caption{Results of the principal component analysis of the experimental images. A combination of the initial state of the Hele-Shaw cell (not shown) and an image at a given time (left) is decomposed into a low rank matrix (center) and a sparse matrix (right). The sparse matrix displays the fingering pattern with significantly less noise than the original image.}
    \label{fig:pcp}
\end{figure}

The light intensity of the experimental images reflected the variable gap width of the heterogeneous Hele-Shaw cells, which prevented the conversion of light intensity to CO$_{2}$ concentration. To remove the effect of the engraved pattern from the images, we performed a robust principal component analysis on the experimental images using the principal component pursuit (pcp) algorithm \cite{Candes2011}. Principal component analysis considers that a matrix $M$ is composed of a low rank matrix plus a sparse matrix. The pcp method finds the low rank and sparse matrices using a minimization algorithm. In our case, $M$ is the matrix resulting of stacking the initial state of the cell with an image at a given time. An example of the image processing can be seen in Figure~\ref{fig:pcp}. After the decomposition, the sparse matrix contains a less noisy picture of fingering patterns more suitable for the instability analysis.
%
%
\subsection{Numerical simulations}
Under the assumptions of incompressible fluid and the validity of the Boussinesq approximation, the governing equations for variable-density single-phase flow in a Hele-Shaw cell are \cite{riaz_2006}:
\begin{align} \label{eq:flow}
  \nabla \cdot \mathbf{u} &= 0
\end{align}
\begin{align}
  \label{eq:Darcy}
  \mathbf{u} &= -\frac{k}{\mu} (\nabla p - \rho \mathbf{g})  
\end{align}
\begin{align}
  \label{eq:transport}
  \frac{\partial c}{\partial t} = - \mathbf{u} \cdot \nabla c + \nabla \cdot \left(\mathbf{D} \nabla c\right)
\end{align}
where $\mathbf{u}$ is the Darcy velocity, $c$ is a normalized CO$_{2}$ concentration, $\rho$ a linear function of concentration
\begin{align}
  \label{eq:density}
  \rho = \rho_0 + \Delta \rho c,
\end{align}
with $\Delta \rho$ a function of the partial CO$_{2}$ pressure \pco\ given by~\eqref{eq:delta_rho} and $\mathbf{D}$ is the dispersion tensor defined as \cite{Scheidegger1961}
\begin{align}
  \label{eq:DispTensor}
  \mathbf{D} = D_{m} \mathbf{I} + \alpha_{T} ||u||\mathbf{I} + (\alpha_{L} - \alpha_{T}) \frac{\mathbf{u}\otimes \mathbf{u}}{||u||},
\end{align}
with $\mathbf{I}$ being the identity matrix, $D_{m}$ the diffusion coefficient and $\alpha_{L}$, $\alpha_{T}$ the longitudinal and transverse dispersion coefficients, respectively.

The dispersion tensor takes into account the mixing induced by velocity variations across the gap thickness. The resulting longitudinal Taylor dispersion can be estimated as \cite{Detwiler2000} 
\begin{equation}\label{eq:DL}
    D_{L} = \frac{Pe^2}{210}D_{m},
\end{equation}
where $Pe = u_{b} E/D_{m}$ is the Péclet number computed using the buoyancy velocity $u_{b}$, gap thickness $E$ and diffusion $D_{m}$. The longitudinal dispersion coefficient $\alpha_{L} = D_{L}/u_{b}$ is then calculated as
\begin{equation}\label{eq:alphaL}
    \alpha_{L} = \frac{u_{b}^{2}E^{2}}{210 D_{m}}.
  \end{equation}
The transverse dispersion coefficient has been observed to be smaller than $\alpha_{L}$ by a factor of 5--20 \cite{Freeze1979}. Here $\alpha_{T}$ is taken as $\alpha_{T} = \alpha_{L}/10$.

For the gap thickness and CO$_{2}$ pressure considered, $1 < Pe <20 $ (see Table~\ref{tab:params}). For most of the considered cases, dispersion was smaller than $D_{m}$.  Only for the cases with a mean gap of 500 $\mu$m and \pco $\ge 60\%$, dispersion was comparable or slightly bigger than $D_{m}$.

Equations (\ref{eq:flow} -- \ref{eq:transport}) were solved in a $L \times H$ rectangular domain. All boundaries are no flow, zero concentration gradient boundaries except for the top one where concentration $c=1$ is prescribed. An error function
\begin{equation}
  \label{eq:c_ini}
  1 - \text{erf}{\left(\frac{H - z}{2 \sqrt{D_{m}t_{0}}}\right)} + a \xi(z) \left[\Theta(z - H + \delta) - \Theta(z - H)\right] ,
\end{equation}
where $\Theta(z)$ is the Heaviside function, $\xi$ a uniformly distributed variable between 0 and 1, $\delta = 0.001$ and $a = 0.01$  are the width and amplitude of the perturbation, respectively, and $t_{0}$ is taken equal to 30~s. Numerical simulations were performed using the variable-density flow and transport solver \texttt{rhoDarcyFoam} belonging to the  open-source computational framework SECUReFoam \cite{Icardi2023}, based on the finite-volume library OpenFOAM \cite{Weller1998}.
%
%
\section{Results and discussions}
This section presents the numerical and the experimental results obtained. We compare the homogeneous and the heterogeneous experiments analyzing the fingers shape, the wavenumber, the amplitude of the instability and its growth rate.


%
%
\subsection{Fingers morphology}
%

\begin{figure}
    \centering
    \includegraphics[width=\textwidth]{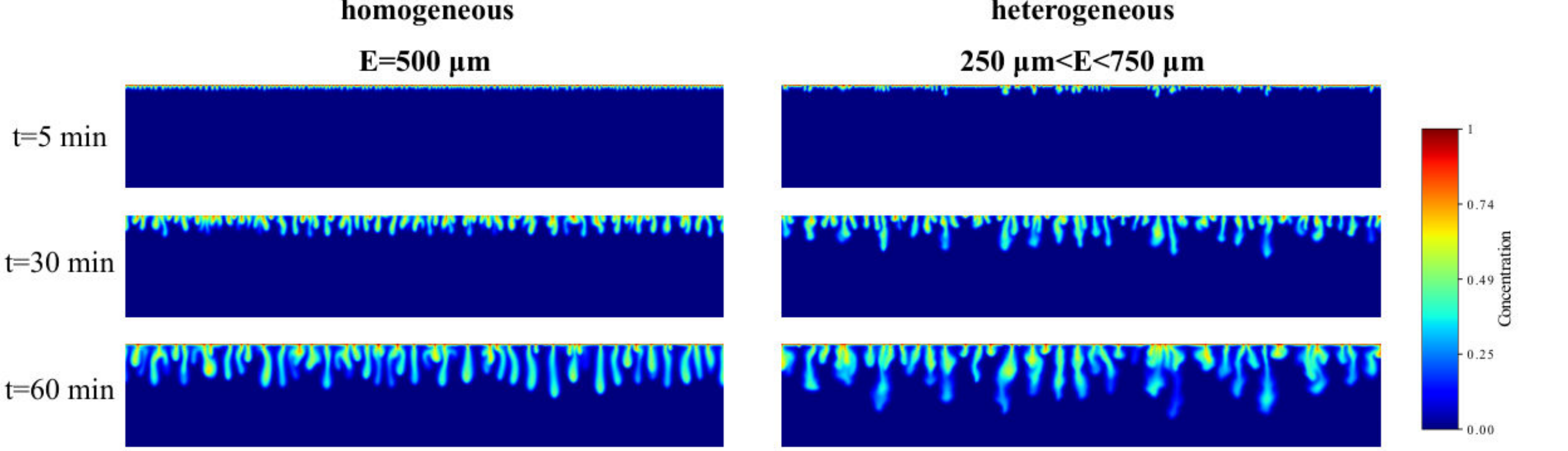}
    \caption{Numerical fingering patterns for the homogeneous ($E=500$ $\mu$m, left) and heterogeneous ($250$ $\mu$m $<E<750$ $\mu$m, right) cases for $P_{\text{CO}_{2}}=85\%$ at 5, 30 and 60 minutes after the onset of convection. Colors show the normalized concentrations.}\label{fig:fingers_num_500}
 \end{figure}

Experimental homogeneous and heterogeneous fingering patterns are noticeably different. In the heterogeneous case, fingers appear distorted and dispersive in comparison to their smoother homogeneous counterparts (Figure~\ref{fig:fingers_exp_500} and figures S1 -- S18 in the supplementary material \cite{supmat}). The same features are observable in the numerical simulations (Figure~\ref{fig:fingers_num_500}). Moreover, the fingers look thicker in the numerical heterogeneous case compared to their homogeneous counterparts. This fingering pattern is in agreement with the classification of Farajzadeh \emph{et al.} (20110) \cite{article_Farajzadeh} based on the Dykstra-Parson coefficient $V_{DP}$ \cite{Dykstra1950, Jensen1997}. Our heterogeneity field  has $V_{DP} \approx 0.5$, which place our system in the dispersive regime fingering pattern. Similar patterns can be obtained for other heterogeneity models as, for example, impervious barriers smaller that the size of the finger size \cite{Green2014}.

Additionally, in the heterogeneous case, fingers exhibit a faster growth leading them to reach the bottom of the medium more rapidly than the homogeneous case. The frequency of finger merging is also higher in heterogeneous cases due to the presence of zones of higher and lower permeability. This contrasts with homogeneous cases, where interactions between fingers are less pronounced, resulting in a reduced merging frequency, a trend consistent in numerical and experimental observations. 
\begin{figure}
    \centering
    \includegraphics[width=\textwidth]{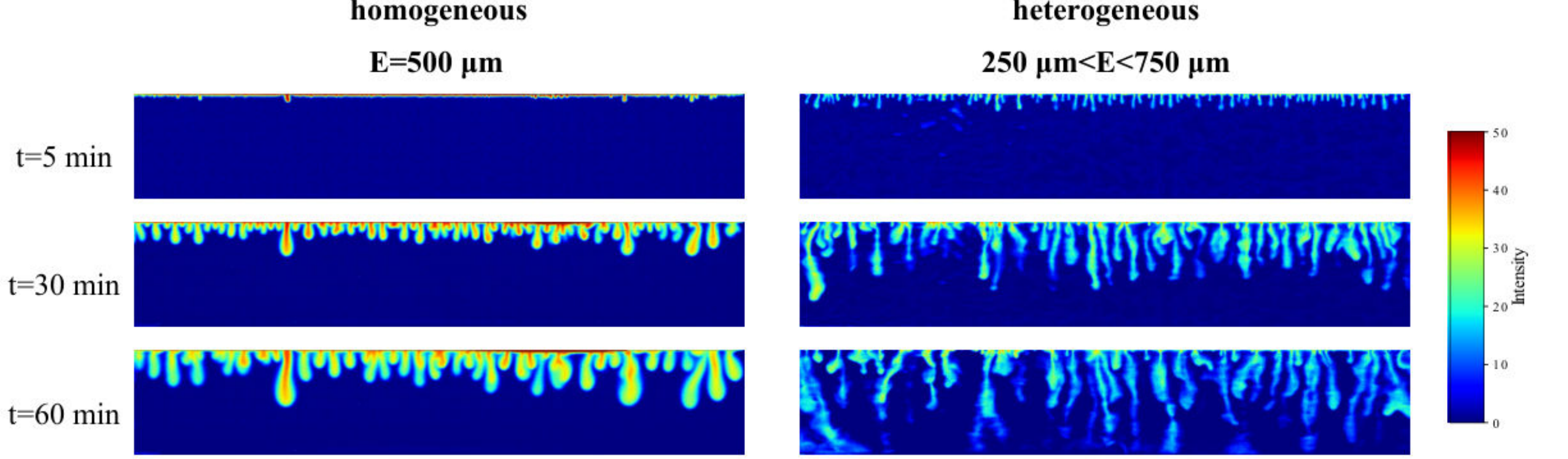}
    \caption{Experimental fingering patterns for the homogeneous ($E=500$ $\mu$m, left) and heterogeneous ($250$ $\mu$m $<E<750$ $\mu$m, right) cases for $P_{\text{CO}_{2}}=85\%$ at 5, 30 and 60 minutes after the onset of convection. Color maps light intensity.} \label{fig:fingers_exp_500}
 \end{figure}
%
%
\subsection{Amplitude of the instability}
The amplitude of the instability is quantified by calculating the standard deviation of the front's position $Z_{F}$
\begin{align}\label{eq:amplitude}
    A^{2}(t) = \overline{Z_{F}(x,t)^{2}  - \overline{Z_{F}(t)}^{2}},
\end{align}
where the overline indicates an average in the horizontal direction. In the numerical results, $Z_{F}$ is determined by  0.1 concentration contour. The experimental images were converted to binary images using a threshold before extracting the contour of the fingering pattern. This process effectively removes the effect of the noise caused by the variable gap width and delineates the boundary where the concentration reaches a specified level.

\begin{figure}
    \centering
    \includegraphics[width=\textwidth]{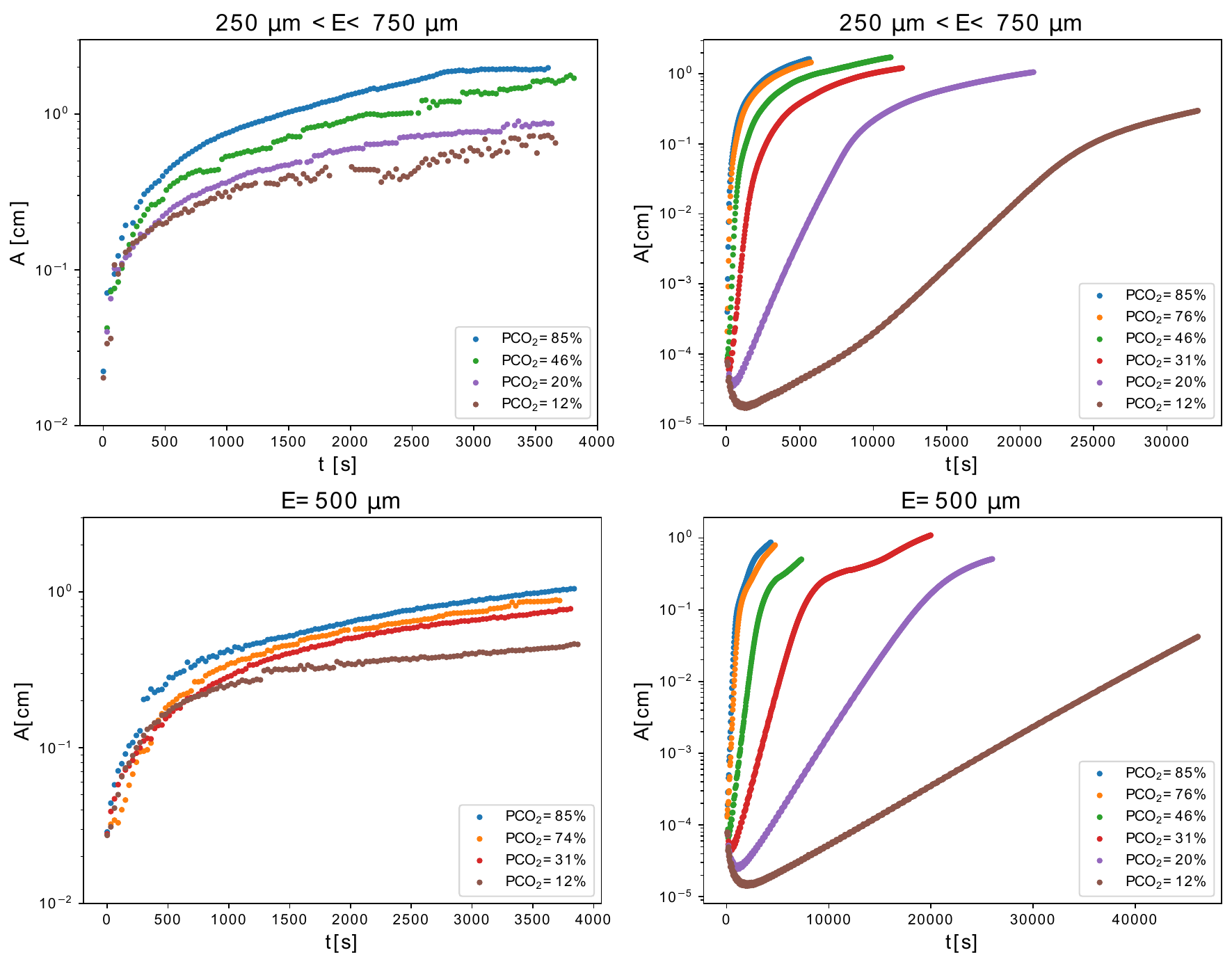}
    \caption{Experimental (left) and numerical (right) instability amplitude versus time from the start of the injection for the $250$ $\mu$m $<E<750$ $\mu$m and the homogeneous case $E= 500$ $\mu$ m obtained experimentally and numerically (for the other cases see figure S20 in the supplementary material \cite{supmat}).}\label{fig:amplitude}
\end{figure}

Figure~\ref{fig:amplitude} shows the amplitude evolution for the heterogeneous case $250$ $\mu$m $<E<750$ $\mu$m and the homogeneous case $E=500$ $\mu$m obtained experimentally and numerically. We observe that heterogeneity increases the amplitude of the front, being approximately twice as large as the homogeneous case. The amplitude is proportional to \pco\ and gap thickness (for different gap cases see figure S19 in the supplementary material \cite{supmat}).

Numerical results display longer times for the development of the instability. However, once the instability develops, similar behavior is observed. The amplitude is slightly smaller in the numerical simulations compared to the experimental observations. This can be attributed to the additional small scale heterogeneity introduced by the engraving of the cell and to a weaker initial noise in the numerical simulations than in the experiments.

%
%
%
\subsection{Instability growth rate}
\begin{figure}
  \centering
  \includegraphics[width=0.48\textwidth]{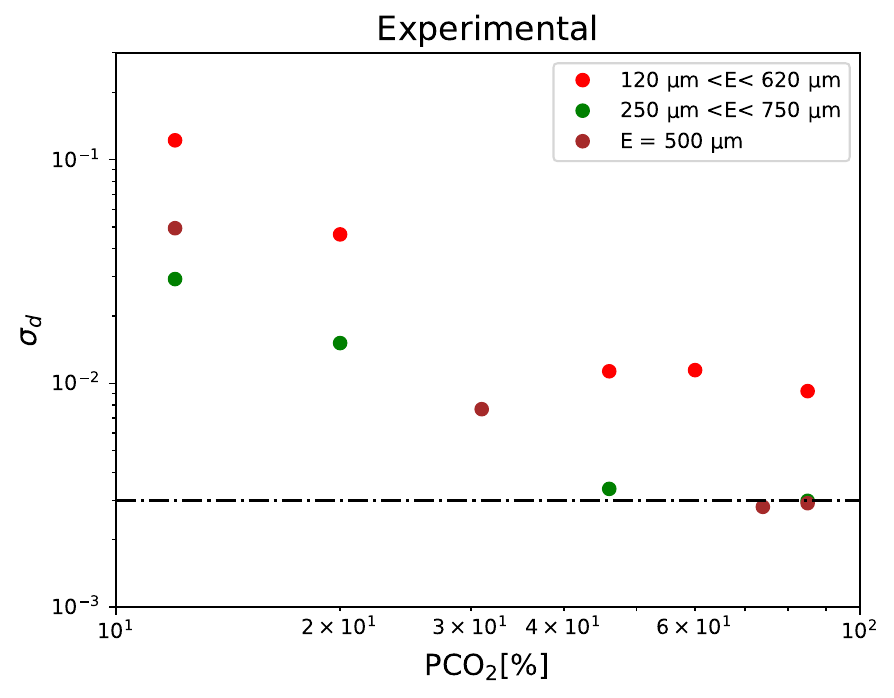}
  \includegraphics[width=0.48\textwidth]{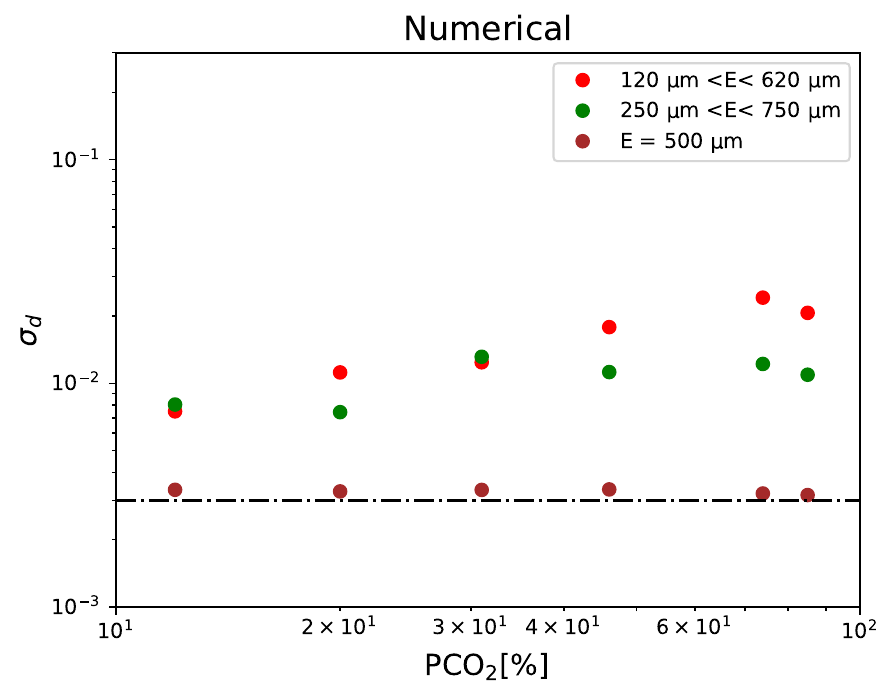}
  \caption{Experimental and numerical instability growth rate, computed from the amplitude at early times vs \pco\ for different gap thicknesses. The dash-dotted line corresponds to the numerical prediction of Elenius \& Johannsen, 2012 \cite{article_elenius} in the Darcy regime.} \label{fig:growth_rate}
\end{figure}

The growth rate of the instability $\sigma$ is derived from the evolution of the amplitude of the front with time (Figure~\ref{fig:amplitude}). The growth rate is defined as the slope observed in the linear phase, which corresponds to the early stage of the instability where it grows steadily \cite{article_elenius}. Since the growth rate scales with $u_{b}^{2}$, $\sigma$ is proportional to \pco\ (and, therefore, $Ra$). To compare the different cases, we define the dimensionless growth rate as
\begin{equation}\label{eq:sigma_d}
  \sigma_{d} = \frac{\sigma D}{u^{2}_{b}}.
\end{equation}
Figure~\ref{fig:growth_rate} shows the numerical and experimental results for $\sigma_{d}$. To account for the uncertainty in the definition of the linear regime, the growth rate was estimated by shifting the beginning and end of the time interval by $\pm 2$ minutes. The numerical growth rate has a relative standard error (RSE) between 0.1\% and 18.6\% with a mean RSE of 8.8\%. The experimental growth rate has a RSE between 3.5\% and 21.5\% with a mean of 3.9\%. In both cases, the biggest error corresponds to the heterogeneous case with lowest \pco\ and the smallest error to the homogeneous case.

The homogeneous numerical simulations closely match the numerical prediction put forth by Elenius and Johannsen (2012) \cite{article_elenius} and previously observed experimentally in low $Da$ Hele-Shaw cells \cite{article_vreme}. The growth rates in the heterogeneous scenarios are roughly three times greater than those of the homogeneous cases with no significant dependence on \pco\ or gap width. The same behavior is observed in the experimental results. The agreement between numerical and experimental results is better for the high values of \pco. This is due to the difficulty to filter the noise for the cases with low CO$_{2}$ concentration and small pH indicator color contrast. In real aquifers, the heterogeneity is usually much larger than chosen in this study. It is thus possible that the instability develops much faster in real conditions than predicted by the classical models without heterogeneity.
%
%
\subsection{Instability wavenumber}
We analyze the instability in terms of most unstable mode using the wavenumber of the instability \cite{article_vreme}. The dimensionless wavenumber is defined as
\begin{align}\label{eq:kd}
 k_{d} = \frac{2\pi N_{f} D}{u_{b}L},
\end{align}
where $N_{f}$ is the number of fingers. The number of fingers is counted by identifying the peaks in a region of interest (ROI) located at the top 5\% of the vertical axis. Gaussian smoothing is applied to the data to reduce the noise and improve the visibility of peaks. Local maxima are detected using a maximum filter, which highlights points that are higher than their neighbors. A threshold, set to the 95th percentile of the smoothed data within the ROI, is used to filter out minor fluctuations and retain significant peaks. The peaks organize themselves in vertical lines that delimit the fingers' centers. Only the highest point in each vertical line is retained to ensure that it corresponds to a finger.  The correspondence between the identified peaks and the fingers is finally verified visually by overlaying the peaks on the smoothed concentration field. An estimation of the variability of the wavenumber was obtained by calculating $k_{d}$ using the distance between each pair of fingers. The numerical wavenumber has a RSE between 1.8\% and 12.8\% with a mean RSE of 5.5\% . The experimental wavenumber has a RSE between 2.8\% and 8.9\% with a mean of 5.4\%. Again the highest errors correspond to the lowest \pco\ heterogeneous cases.

\begin{figure}
    \centering
    \includegraphics[height=10cm]{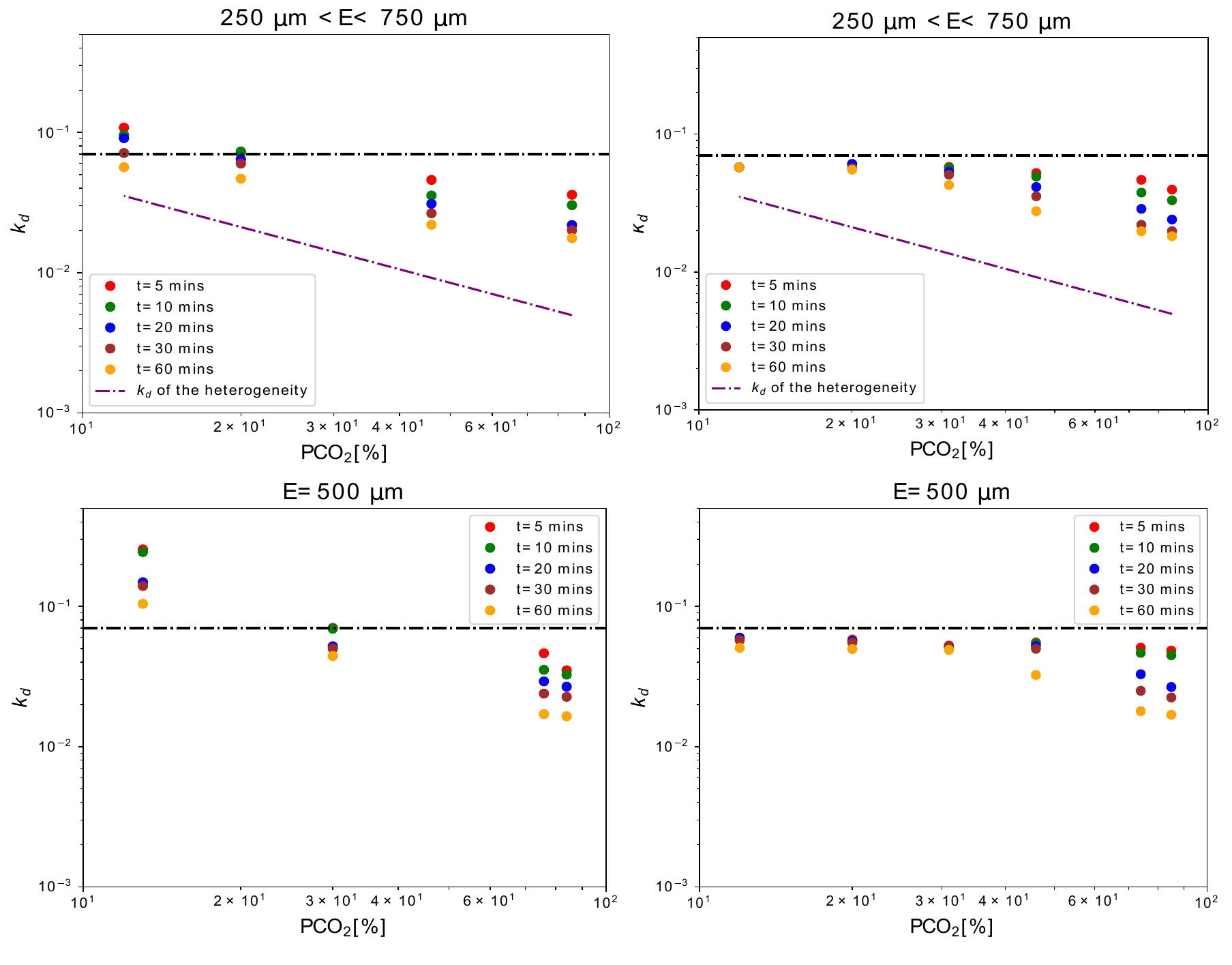}
    \caption{Experimental (left) and numerical (right) dimensionless wavenumber vs \pco\ for $E = 500$ $\mu$m. The dash-dotted line corresponds to the theoretical prediction (of the critical wavenumber $k_{th}$) obtained by Riaz et al., 2006 \cite{riaz_2006}. The dashed line corresponds to the heterogeneity wavenumber.}\label{fig:wavenumber}
\end{figure}

Figure~\ref{fig:wavenumber} shows $k_{d}$ for the cases $250$ $\mu$m $<E<750$ $\mu$m and $E= 500$ $\mu$ at selected times. Time is measured relative to the time for the onset of convection for the numerical results.  Experimentally, convection starts immediately after the stop of the CO$_{2}$ injection, while numerically convection is delayed differently depending on \pco. The numerical onset time of the instability is taken as the time for the  minimum flux, calculated as the diffusive flux on the top boundary (see figure S22 and S23 in supplementary material \cite{supmat}).

The values obtained for the wavenumber in the homogeneous case at early times closely match the theoretical value $k_{th} = 0.07$ obtained by Riaz et al. (2006) \cite{riaz_2006}). The agreement is particularly good in the numerical simulations. We observe a decrease in $k_{d}$ with time, caused by the reduction of $N_{f}$ as fingers merge.

Similarly to the observations by Li \emph{et al.} (2019) \cite{article_Li2020}, a comparison between the homogeneous and heterogeneous cases reveals that, both numerically and experimentally, $k_{d}$ does not change significantly with heterogeneity particularly with lower \pco. This is probably due to a locking of the instability to the wavelength of the forcing due to the heterogeneity. Indeed, the theoretical dimensionless wavenumber $k_{th}$ of the instability is close to the dimensionless wavenumber of the heterogeneity ($2 \pi D/(u_{b} \lambda_x)$, purple dash-dotted line in Figure~\ref{fig:wavenumber}) at small \pco.

The difference between the two gap widths considered is smaller than the effect of \pco\ or heterogeneity (see figure S19 in supplementary material \cite{supmat} for the $E=370 \mu$m homogeneous and heterogeneous cases). However, for small \pco, the wavenumber tends to be a little higher for $E=500 \mu$m than for $E=370 \mu$m, suggesting that finger merging is more efficient with larger gap thicknesses.  
%
%
\subsection{Interaction between heterogeneity and fingering instability}
To analyze the effect of the heterogeneity on the evolution of the instability, we analyze the overall fingering patterns by means of the autocorrelation function defined as
\begin{align}
  \label{eq:acf2d}
  \text{ACF}_{c}(x, z) = {\cal{F}}^{-1}\left\{\left|{\cal{F}}\left\{c(x,z)\right\}\right|^{2}\right\},
\end{align}
where $c(x,z)$ is the function representing the fingering pattern and ${\cal F}$ is the two-dimensional Fourier transform. The shape of ACF indicates the presence of periodic structures. The size of the periodic structure can be determined by fitting $1 - \text{ACF}(x>0, H/2)$ to a Gaussian variogram \cite{Webster2007-bz}
\begin{align}
  \label{eq:variogram}
  \gamma(x) = s_0 \left[ 1 - e^{-\left(x/a\right)^{2}} \right],
\end{align}
where $s_{0}$ is the sill and $a$ is the range of the variogram. We define the effective range of the variogram as the distance $r$ for which $\gamma(r) = 0.95 s_0$. For the Gaussian variogram $r= \sqrt{3}a$. It is, therefore, representative of the correlation length of the pattern and, thus, of the finger width. We limit ourselves to the analysis of the numerical results because the Fourier spectrum of the experimental images was too noisy even after being processed with the pcp method.

\begin{figure}
\centering
\includegraphics[width=0.8\textwidth]{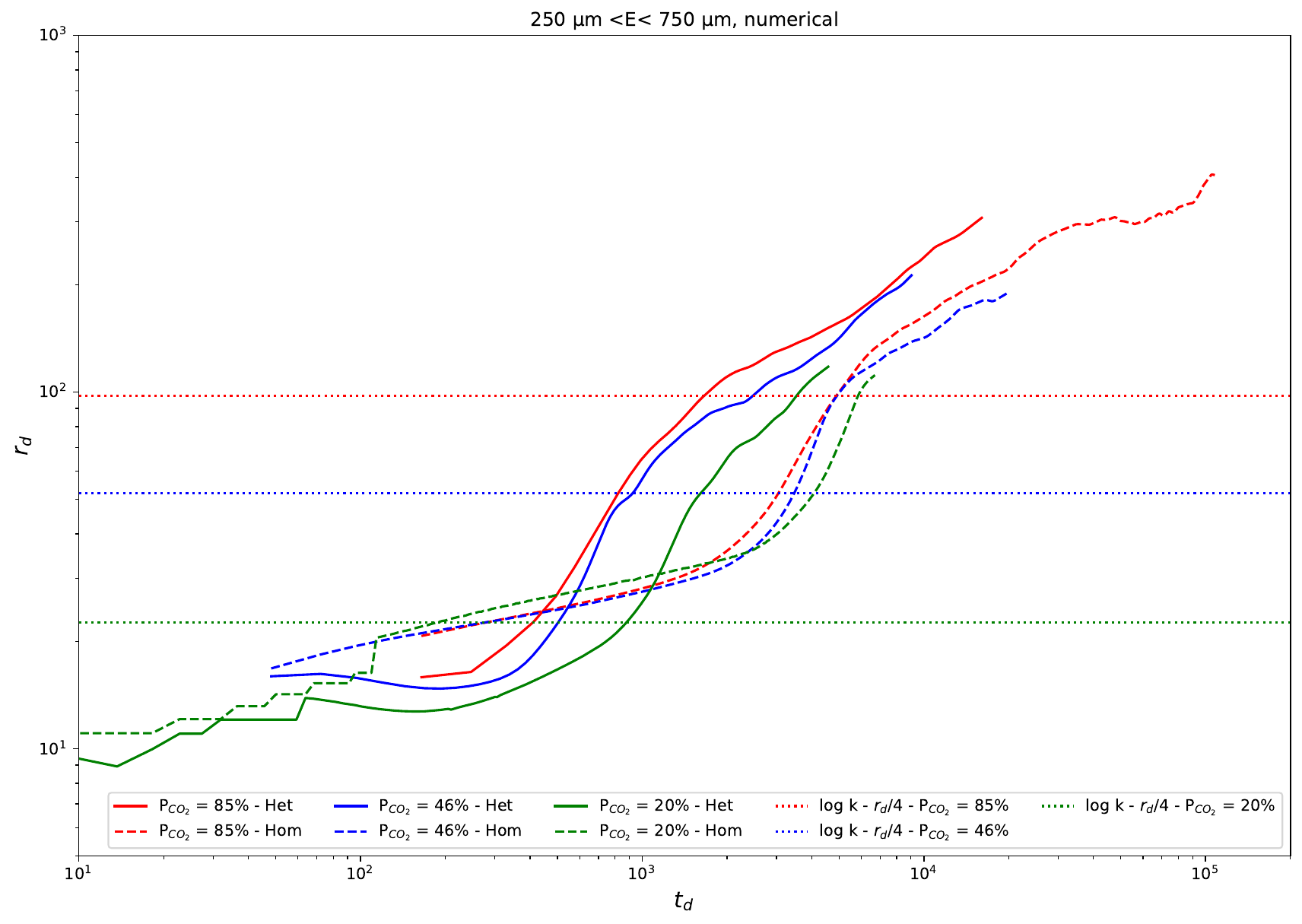}
    \caption{Comparison of the dimensionless numerical patterns range and dimensionless wavelength versus time for the cases $E=500 \mu$m and $250$ $\mu$m $<E<750$ $\mu$m.} 
    \label{fig:ranges_num500}
\end{figure}

Figure \ref{fig:ranges_num500} presents the evolution of the dimensionless effective range $r_{d} = r u_{b}/D_{m}$ versus the dimensionless time $t_{d} = t u_{b}^{2}/D_{m}$ of the patterns together with the dimensionless effective ranges of the permeability field (see figure S21 in the supplementary material \cite{supmat} for the other case with $120$ $\mu$m $<E<620$ $\mu$m). Initially, high \pco\ cases have a smaller $r_{d}$ than low \pco\ cases. This corresponds to the small fingers observed with high \pco. However, the higher \pco\ is, the earlier and faster $r_{d}$ grows. At late times, high \pco\ cases have a higher $r_{d}$. The range in heterogeneous cases increases at a faster rate compared to the homogeneous cases. The range is always larger for the heterogeneous cases, which supports the conclusion that heterogeneous fingers are thicker and less spaced apart than their homogeneous counterparts. It is also consistent with the smaller dimensionless wavenumber observed in the heterogeneous cases. We also observe two regimes in the range evolution. First, a sudden growth after the onset of convection and then a change in the slope around one fourth of the range of the log-permeability field (horizontal dotted line in Figure \ref{fig:ranges_num500}). When the ranges of the heterogeneity and the instability are comparable, the instability-range growth slightly diminishes, suggesting that the relative size between the fingers and heterogeneity affects the pattern size.
%
%
\section{Conclusions}
We studied CO$_{2}$ dissolution in a heterogeneous Hele-Shaw cell numerically and experimentally. Fingers morphology shows that patterns exhibit more dispersion and distortion in heterogeneous scenarios compared to their smoother homogeneous counterparts. The evolution of the size of the fingering patterns given by the autocorrelation function showed that heterogeneity accelerates the growth of the fingering patterns with respect to the homogeneous cases. This is confirmed by the growth rate obtained from the amplitude of the instability. Low \pco\ cases seem to be less affected by the heterogeneity as reflected by the behavior of the instability wavelength. This suggests an interaction between the scales of the heterogeneity and the instability that affects the pattern evolution, which deserves further investigation. In conclusion, heterogeneity amplifies the convective instability's amplitude and accelerates the growth rate. This implies that heterogeneous porous media enhance secure CO$_{2}$ storage by enhancing CO$_{2}$ dissolution, thereby mitigating leakage risks.
%
%
\begin{acknowledgments}
This project has received funding from the European Union’s Horizon 2020 research and innovation program under the Marie Skłodowska-Curie grant agreement N°956457. RB and JJH acknowledge the support of the MICIU/AEI/10.13039/501100011033 and the European Union NextGenerationEU/PRTR through the grant ESFERA CNS2023-144134.
\end{acknowledgments}
%
%
%
%
%
\end{document}